\documentclass[aps,prl,superscriptaddress]{revtex4}

\usepackage{graphicx}
\usepackage{ifthen}

\newif\ifpdf\ifx\pdfoutput\undefined\pdffalse\else\pdfoutput=1\pdftrue\fi

\newcommand{\mygraphics}[2][]{\centering \includegraphics[#1]{#2}}
\newcommand{\eqa}{\begin{eqnarray}} 
\newcommand{\eqe}{\end{eqnarray}} 
\newcommand{\8}{\begin{equation}}
\newcommand{\9}{\end{equation}}

\begin{document}

\title{Quantum Eavesdropping without Interception: An Attack Exploiting the Dead Time of Single Photon Detectors}

\author{Henning Weier}
\email[]{henning.weier@lmu.de}
\affiliation{Fakult\"at f\"ur Physik, Ludwig-Maximilians-Universit\"at, 80799 Munich, Germany}
\affiliation{qutools GmbH, 80539 Munich, Germany}

\author{Harald Krauss}
\affiliation{Fakult\"at f\"ur Physik, Ludwig-Maximilians-Universit\"at, 80799 Munich, Germany}

\author{Markus Rau}
\affiliation{Fakult\"at f\"ur Physik, Ludwig-Maximilians-Universit\"at, 80799 Munich, Germany}

\author{Martin F\"urst}
\affiliation{Fakult\"at f\"ur Physik, Ludwig-Maximilians-Universit\"at, 80799 Munich, Germany}
\affiliation{qutools GmbH, 80539 Munich, Germany}

\author{Sebastian Nauerth}
\affiliation{Fakult\"at f\"ur Physik, Ludwig-Maximilians-Universit\"at, 80799 Munich, Germany}
\affiliation{qutools GmbH, 80539 Munich, Germany}

\author{Harald Weinfurter}
\affiliation{Fakult\"at f\"ur Physik, Ludwig-Maximilians-Universit\"at, 80799 Munich, Germany}
\affiliation{Max-Planck-Institut f\"ur Quantenoptik, 85748 Garching, Germany}

\date{\today}

\begin{abstract}

The security of quantum key distribution (QKD) can easily be obscured if the eavesdropper can utilize technical imperfections of the actual implementation. Here we describe and experimentally demonstrate a very simple but highly effective attack which even does not need to intercept the quantum channel at all. Only by exploiting the dead time effect of single photon detectors the eavesdropper is able to gain (asymptotically) full information about the generated keys without being detected by state-of-the-art QKD protocols. In our experiment, the eavesdropper inferred up to 98.8\% of the key correctly, without increasing the bit error rate between Alice and Bob significantly. Yet, we find an evenly simple and effective countermeasure to inhibit this and similar attacks.

\end{abstract}


\maketitle

The communication of sensitive data has become part of our everyday life resulting in a growing need for mechanisms which ensure secure transmissions of these data. The secrecy of the information transfer can be guaranteed using a classical cryptographic method called one-time-pad. This method enables unconditionally secure communication -- provided that the exchange of the cryptographic key has been perfectly secure. In 1984 Ch. Bennett and G. Brassard showed that this indeed can be achieved using quantum cryptography, or more precisely quantum key distribution (QKD) \cite{bb84,gisin_rev_mod_phys}, an approach which employs non-orthogonal quantum states for encoding information. Over the past years there have been remarkable QKD experiments pushing both the limits in distance and/or key rate \cite{bienfang:04,rosenberg_decoy_fiber,tschmitt,shields_high_rate_qkd} as well as the level of applicability achieving network functionality \cite{SECOQC-demo, DARPA-demo} with first systems for quantum secured communication being commercially available.

Yet, what does "secure" mean? Today there exist security proofs \cite{lo_chau,shor_preskill} showing that the ideal protocol is secure in the sense that any knowledge of an eavesdropper about the key can be quantified and consequently made negligibly small.
However, these proofs rarely specify requirements for QKD hardware and, if they do, real implementations will usually not fully comply with these specifications. This can lead to new types of attacks which are not covered by the proofs and hence won't be revealed by standard security tests. Recently, considerable effort has been made to reveal those potential threats \cite{Lamas-Linares:07,gelles_mor_quantum_space_attacks,lo_quantum_hacking,nauerth_side_channels,lo_phase_remapping_attack} and to find countermeasures against them \cite{ILM,GLLP,dec_hwang,dec_wang1,dec_ma1,dec_ma2,fung_det_eff_proof}.
Many attacks are designed only for very specific systems and/or require sophisticated technology which is not yet (public) state-of-the-art.

In this paper we introduce a novel type of attack which even does not need to intercept the qubits sent over the quantum channel. We show how to utilize an imperfection, which almost all QKD-systems display, namely the fact that common single photon detectors are rendered inactive for a period of time (called dead time) after a detection event. This enables the eavesdropper to unveal the full key without significantly changing the quantum bit error ratio using very simple equipment.
On the one hand we demonstrate that this new attack renders a conventional QKD system absolutely insecure, but, on the other hand, we also provide an effective countermeasure.

There are two characteristic features of nearly all QKD systems implemented so far. The first is the fact that when a SPAD registers a photon, there usually follows a period of time during which it will not be able to detect a second event. This period (called dead time $\tau_D$) can range from less than a nanosecond to some tens of microseconds.
The second feature is the periodic operation: The transmitter emits signals only at well defined times $t_i = i \cdot T$, with period $T$. Consequently, in order to reduce noise originating from intermediate dark counts and scattering events, the receiver accepts a detection event only during a narrow time window ($\Delta_{tw} \ll T$) around $t_i$ -- all events outside these time windows are discarded.

\begin{figure}
\mygraphics[width=8.7cm]{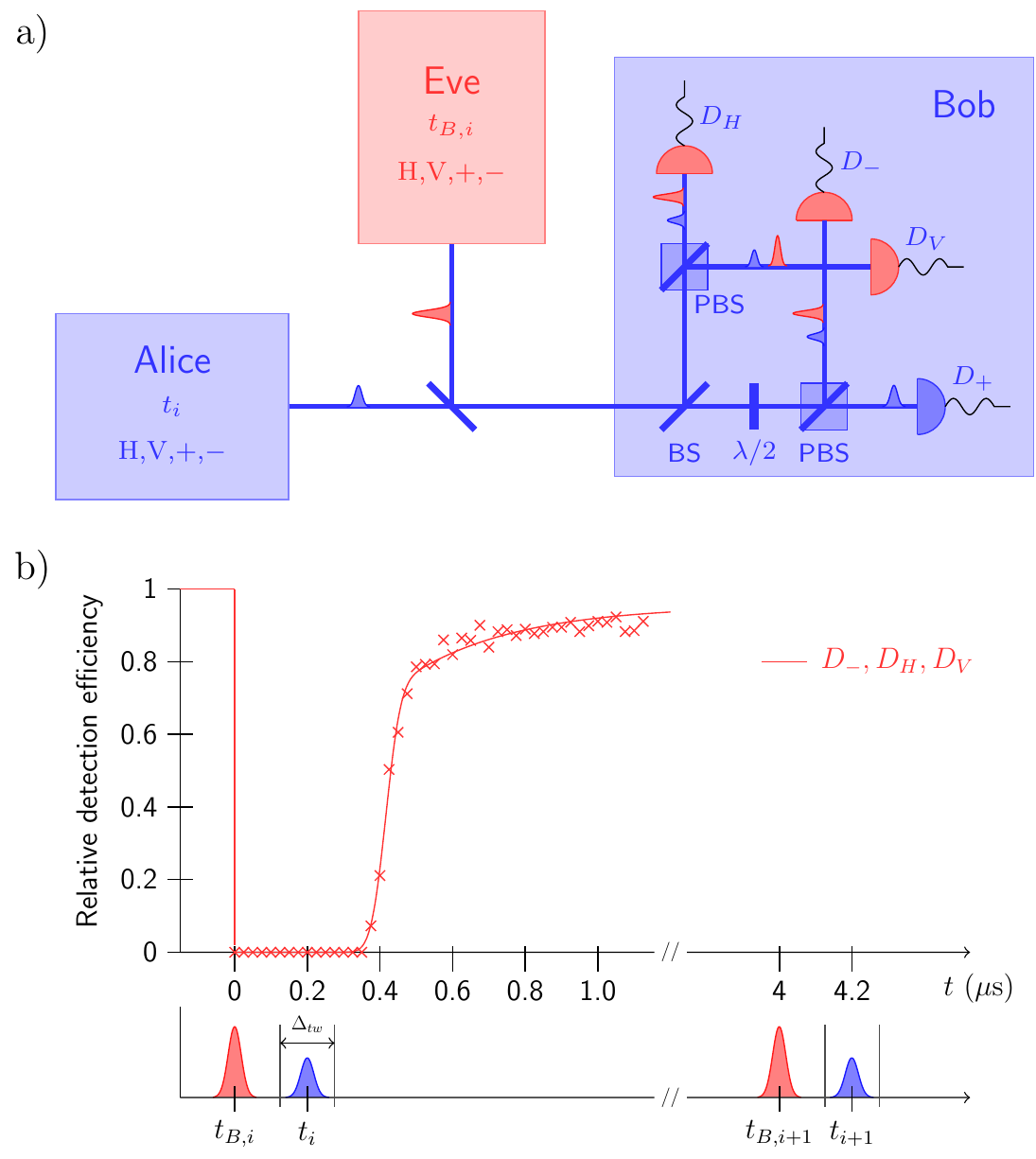}
\caption{a) Schematic drawing of the set-up: The transmitter (Alice) on the left, her pulses marked blue, the eavesdropper (Eve) in the centre, her pulses marked red and the receiver (Bob) on the right. The latter performs a state analysis with four detectors and passive optical components. The module utilizes a first non-polarizing 50/50 beam splitter (BS), which effectively serves as a (quantum) random basis switch, such that if the incident photon is reflected and detected behind a polarizing beam splitter (PBS) it is analyzed in the H/V basis, whereas, if transmitted by the BS and a half-wave plate @~$22.5^\circ$ and detected behind a second PBS it is analyzed in the $\pm45^\circ$ basis. In this example, when Eve sent a light pulse polarized in one of the four polarizations (here: $-45^\circ$), up to three of Bob's detectors (in this example $H,-45^\circ,V$, red) become inactive with a certain probability. Eve thus knows that Alice's signal pulse can only be detected in the remaining detector (here $+45^\circ$) and can infer the respective key bit.
b) Relative efficiency for a second detection event after a detection at $t=0$ (see Methods) together with the timing of Eve's blinding pulses at $t_{B,i}$ and Alice's signal pulses at $t_i$. (Pulse and gate widths are not to scale, typically $\sim 1$~ns.)
}
\label{fig_blend_prinzip}
\end{figure}

Several attacks have been proposed making use of the dead time of SPADs to enable intercept-resend eavesdropping strategies \cite{makarov_active_quenched,makarov_passive_attack, erlangen_dead_time}. Effectively, for these attacks the eavesdropper employs a sophisticated intercept-resend setup and uses bright light to gain full control over the SPADs of the receiver and to generate detection events equal to his own. The attack demonstrated here is technically much simpler and neither intercepts the quantum channel nor does it require bright light to control the receiver or to spy into this system by some Trojan horse attack. Rather, it utilizes the fact that dead times enable the eavesdropper to manipulate the detection efficiencies for a short time around $t_i$ by blinding some of the installed detectors. Although the eavesdropper does not need to intercept the qubits sent over the quantum channel, she can reveal the full key without being detected.

For our blinding attack, the eavesdropper (Eve) simply sends attenuated light pulses of one of the four polarizations used in the protocol (i.e. H, V, $+45^\circ$ or $-45^\circ$), into the quantum channel at time $t_{B,i}$, shortly (less than $\tau_D$) before $t_i$ and of course still outside of Bob's time window ($\Delta_{tw}/2<t_i-t_{B,i}<\tau_D$). Depending on intensity and polarization of the pulse, Bob's SPADs detect this light with a certain probability, except for the SPAD detecting the orthogonal polarization. Bob then is partially blinded (note, in QKD protocols the events caused by blinding pulses are not taken into account as they are outside of Bob's time window) and if he agrees with Alice on using a particular event in the sifting phase Eve will have significant information about the respective key bit as it could have been detected only by not blinded detectors. Eve can easily tune the intensity of her blinding pulses and thus the information about the key. As it turns out, dim pulses containing only a few photons are sufficient to determine almost all the key (see Methods).

\begin{figure}
\mygraphics[height=8.7cm, angle=270]{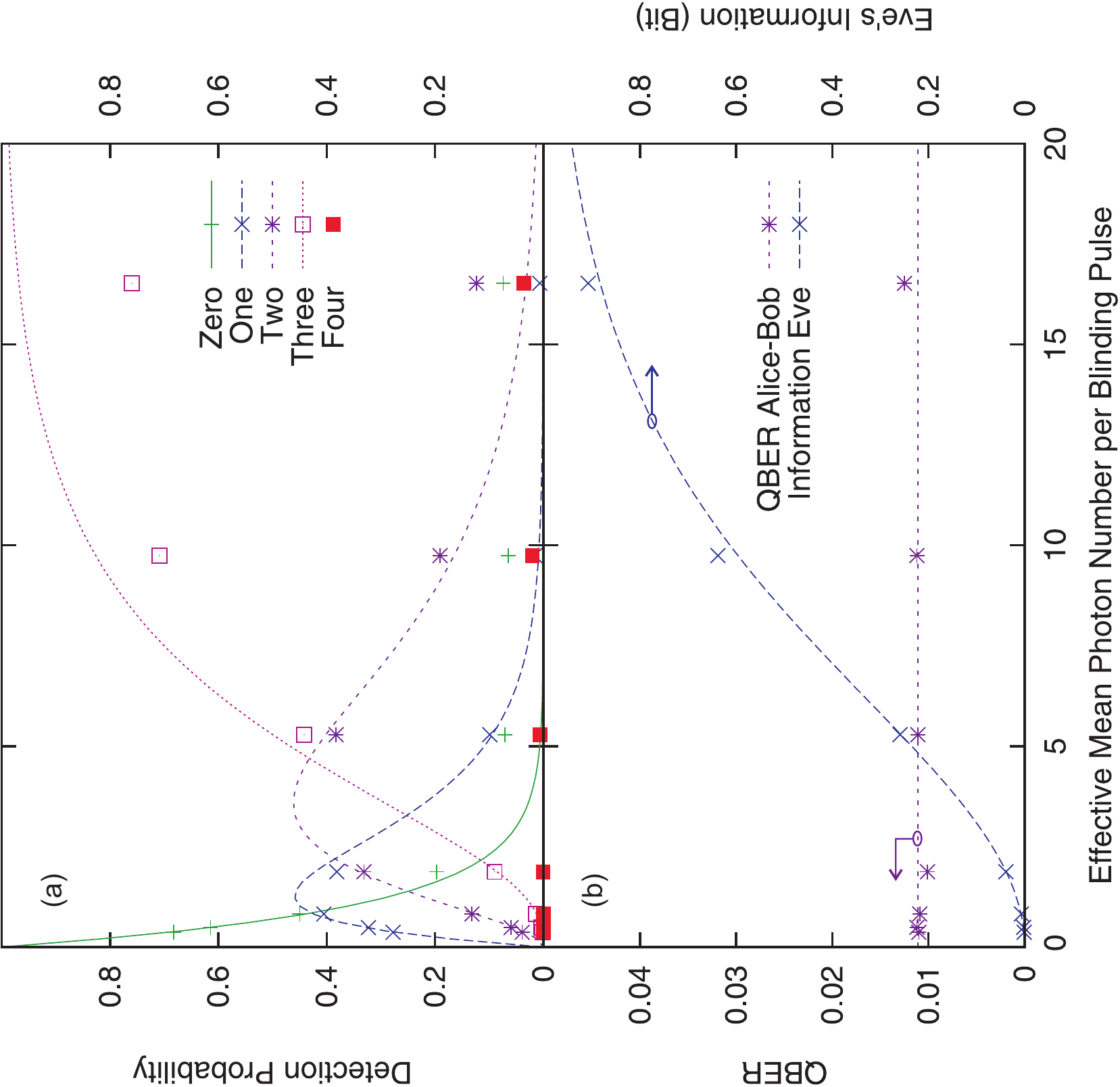}
\caption{
Experimental results of the eavesdropping attack (a) showing the probability for detecting blinding pulse photons in zero, one, two, three or four detectors depending on the effective mean photon number per blinding pulse. (b) shows the quantum bit error ration (QBER) of Alice's and  Bob's key and the information about this key gained by Eve. By using different combinations of neutral density filters, the mean photon number per blinding pulse was changed, while keeping the signal pulse intensity constant. With increasing number of photons per blinding pulse the probability that three detectors are blinded increases rapidly resulting in an information of $>0.9$ bit for about 17 photons, while the QBER does not increase and leaves Alice and Bob ignorant about the attack.
}
\label{fig_blink_theory}
\end{figure}

In the following we describe a model QKD device which employs a BB84 protocol \cite{bb84} with the polarization of single photons or attenuated light pulses encoding the qubits. The general principle can be easily transferred to nearly all other QKD-systems \cite{shields_high_rate_qkd, shields_makarov_bashing}.
We set up a copy of our free space QKD system \cite{pop_dach,SECOQC-demo} in the lab. There, Alice used a four diode transmitter and was connected by a short free space quantum channel to Bob's four SPAD receiver module (Fig. \ref{fig_blend_prinzip}a). Both, transmitter and receiver units where fully computer controlled and ran a real-time BB84 protocol.
Additionally, Eve coupled dim pulses from a transmitter module, similar to the Alice module, into the quantum channel. For easier synchronization with Alice and Bob, control signals for the Eve module were obtained from Alice's controller.

The timing was set such that Alice's signal pulses were sent with a period of  $T=4~\mu$s (long enough to allow the SPADs to recover with a high probability between two consecutive signal pulses and to guarantee unbiased detections).
In accordance with the timing conditions of our setup ($\tau_D \sim 2~\mu$s, $\Delta_{tw} = 5$~ns) the randomly polarized blinding pulses were sent $200$~ns before the signal pulses. The mean photon number $\mu_B^{eff}:=\eta_{B}\mu_B$ of the blinding pulses was set by inserting different combinations of neutral density filters. Here, $\mu_B^{eff}$  constitutes the mean photon number per pulse Eve would have to send into an ideal detector module (regarding transmission and detector efficiency) built like depicted in Fig. 1. $\eta_B$ describes the transmission from Eve to Bob. Uniform, but non-unity transmission and detector efficiencies can be included here. For receiver modules with active basis switching (two detectors), only half of the blinding pulse intensity is necessary.

Eve's key was deduced solely from the knowledge about the setting of her blinding pulses and from eavesdropping the classical communication between Alice and Bob.
To demonstrate the efficiency of this attack, Eve applied different blinding intensities during regular runs of the BB84 protocol between Alice and Bob. The calculated blinding pulse detection probabilities (Fig. 2a) are in good agreement with the experimental data particularly for low blinding pulse intensities $\mu_B^{eff}$. For higher values of $\mu_B^{eff}$, the predictions differ mainly because of a higher number of background events due to increased spontaneous emissions from Eve's laser diodes between the pulses. These, too, can render the detectors inactive and thus reduce the probability for multi-photon detection events due to blinding pulses.
In our experiment no hardware gating was used, i.e. the SPADs are in principle always active. Yet, as gate times are device information we can safely assume that the eavesdropper knows about the timing of the detector efficiency relative to the signal pulses and will act accordingly.

\begin{figure}[t]
\mygraphics[width=8.7cm]{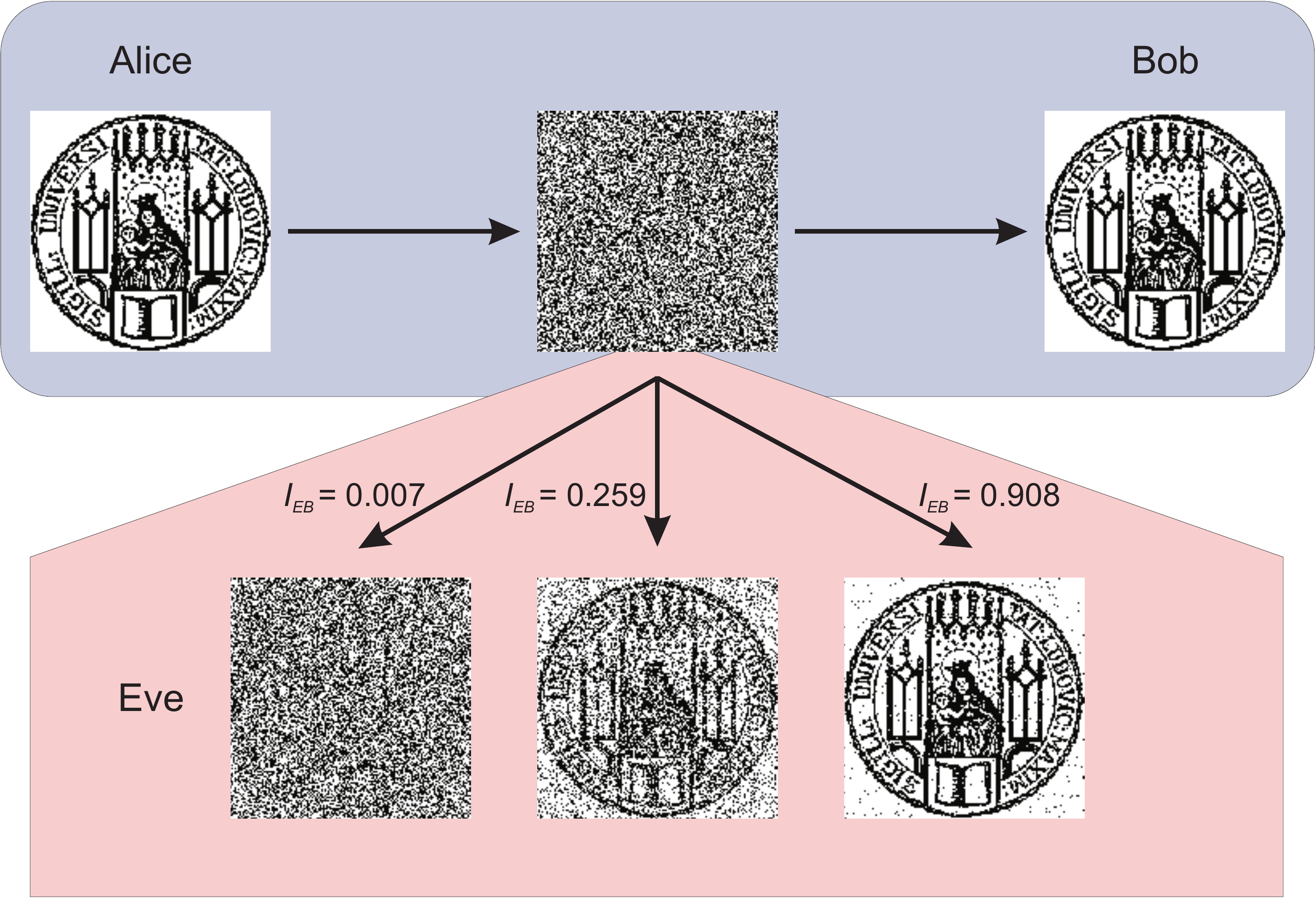}
\caption{\label{fig_lmu_enc}Application of the keys obtained in the experiment. Alice uses her sifted key to encrypt the original image with a one-time pad (top left corner) and sends the ciphertext (top center) to Bob, who uses his error corrected key to decrypt the image (top right corner). The three images below are decrypted using Eve's educated guess of the sifted key, for blinding pulses with a mean photon number of $\mu_B^{eff} = 0.83$, $\mu_B^{eff}=5.29$ and $\mu_B^{eff}=16.52$, respectively.}
\end{figure}

As expected, for low blinding pulse intensities, Eve's attack has only a low probability of success and her key is hardly correlated with the sifted key between Alice and Bob. Yet, by slightly increasing the power of her pulses, the match between the keys rises rapidly. The maximum observed overlap between Bob's and Eve's sifted keys was as high as $98.83 ~\%$ at a blinding pulse mean photon number of only $\mu_B^{eff} = 16.52$, corresponding to a mutual information $I_{EB}=0.908$~Bit (Fig. 2b).
Figure \ref{fig_lmu_enc} visualizes the success of Eve's attack, who easily recovers the emblem of the University of Munich from the one-time pad encrypted cipher.

In conclusion, we have demonstrated a powerful and successful attack that threatens many state-of-the-art QKD systems. By inserting blinding pulses into the quantum channel and eavesdropping on the classical communication only, an adversary is able to gain almost full information about the sifted key without being detected. The potential of this attack is especially high because of its simplicity. The eavesdropper does not need to intercept the quantum channel and does not need to measure the low light level photonic signals emitted by Alice.

Fortunately, the defence against this blinding attack is as simple. Evidence could be obtained, if Bob analyzes the detection events not only during the short time windows. However, by cleverly employing (several) blinding pulses at random times during this interval, Eve would simulate background noise and her attack still could remain unnoticed. A better strategy would monitor the status of the SPADs. This can be deduced from the bias voltage at the SPADs such that it is guaranteed that the detection efficiency is at a nominal level. Now, if we use only those detection events for the key generation where {\em all} detectors were active, the blinding attack and all other currently proposed dead time attacks \cite{makarov_active_quenched,makarov_passive_attack,erlangen_dead_time} become ineffective. This scheme also avoids possible problems due to saturation effects in ultra high rate QKD set-ups \cite{rogers_deadtimes,Xu:06}, thereby establishing significant trust in this quantum secure photonic communication method.

\begin{table}
\begin{tabular}{c|c|c|c}
\hline \hline
$\mu^{eff}_B$& QBER Alice-Bob / \% & QBER Bob- Eve / \% & $I_{EB}$\\
\hline
0.37 & $1.10$ & $48.17$ & $0.001$\\
0.49 & $1.12$ & $47.54$ & $0.002$\\
0.83* & $1.09$ & $45.24$ & $0.007$\\
1.88 & $1.01$ & $38.43$ & $ 0.039$\\
5.29* & $1.11$ & $21.00$ & $0.259$\\
9.75 & $1.12$ & $6.91$ & $0.638$\\
16.52* & $1.25$ & $1.17$ & $0.908$\\
\hline \hline
\end{tabular}
\caption{\label{tab_results} Quantum bit error ratios (QBER) and Eve's information for the different blinding pulse intensities used in the experiment. The three datasets marked with a "*" are the ones used to decrypt the secret message shown in Fig. \ref{fig_lmu_enc}. Note, there is only a minor change of the QBER between Alice and Bob making it impossible for them to discover the attack. The sifted key files were each longer than 20~kByte, resulting in a statistical error of the QBER of $<3\%$.}
\end{table}

\appendix*

\section{Dead Time Analysis}
SPADs exhibit a detection efficiency which depends on the overbias voltage applied. After detection, depending on the electronics, it takes some time until the detector regains full efficiency. 
For the characterization of the detector's dead time, we illuminated it by two consecutive faint laser pulses. The delay between the first and second pulse has been varied and the corresponding relative detection efficiency (normalized to the value after 3.5~$\mu$s was recorded (see Fig. 1b). The line is a fit using the function $E(t')=\frac{1}{2}\left(1+\mathrm{erf} \left(\frac{t'-\tau_D}{\tau_2}\right) \right) \cdot \left(1-e^{-\frac{t'}{\tau_3}}\right)$ with $t'$ time after detection and fit parameters $\tau_D$,$\tau_2$ (jitter due to pulse discrimination) and $\tau_3$ (charging time of SPAD capacity). For passive quenching electronics the dead time $\tau_D$ is particularly long ($\approx 400$~ns), whereas it is about 50~ns when using active quenching. Nevertheless, the eavesdropping scheme can be applied equally well.

\section{Detection Probabilities}

\subsection{Blinding pulse detection probabilities}

To estimate Eve's information (Fig. 2) we assume that the delay between two of Bob's signal time slots and also between Eve's blinding pulse and the preceding signal pulse is greater then the average dead time of the SPADs, which itself is longer than a signal time slot. We will further assume the recovery process to be binary (on or off) with a certain dead time and a passive basis choice setup with four detectors (see Fig.~\ref{fig_blend_prinzip}). Active switching systems with two detectors can be analyzed accordingly.

We first calculate the detection probabilities of blinding pulses (coming from Eve) and signal pulses (emitted by Alice). We start with the blinding pulses. Let $P_{p}(\mu_B^{eff})$ and $P_{d}(\mu_B^{eff})$ be the probability that a blinding pulse is recognized in the detector analyzing parallel and diagonal polarization relative to the blinding pulse polarization, respectively. The corresponding detection probability in the orthogonal orientation is negligible. Detection probabilities depend on the blinding pulse intensity expressed as the mean photon number per pulse $\mu_B^{eff}$ coupled into the (ideal) quantum channel. Here we include the coupling efficiency from Eve to Bob.

The probabilities of registering detections in one, two or three of the respective detectors at the same time then are:
\eqa
p_{p} (\mu_B^{eff}) &=& \underbrace{P_{p} (\mu_B^{eff})}_{\mbox{\small parallel detector clicks}} \cdot \underbrace{(1-P_d(\mu_B^{eff})) (1-P_d(\mu_B^{eff}))}_{\mbox{\small both diagonal detectors do not click}}\\
p_d (\mu_B^{eff}) &=& P_d (\mu_B^{eff}) \cdot (1-P_p(\mu_B^{eff})) (1-P_d(\mu_B^{eff})) \qquad \qquad (2\times)\\
p_{pd} (\mu_B^{eff}) &=& P_p (\mu_B^{eff}) \cdot P_d (\mu_B^{eff}) \cdot (1-P_d(\mu_B^{eff})) \qquad \qquad (2\times)\\
p_{dd} (\mu_B^{eff}) &=& P_d^2 (\mu_B^{eff}) \cdot (1-P_p (\mu_B^{eff}))\\
p_{pdd} (\mu_B^{eff}) &=& P_p(\mu_B^{eff}) \cdot P_d^2(\mu_B^{eff})
\eqe

So the probabilities that none, one, two or three detectors fire due to a blinding pulse are (Fig. 2):
\eqa
p_{(0)}(\mu_B^{eff}) &=& (1-P_p (\mu_B^{eff})) \cdot (1-P_d (\mu_B^{eff}))^2 \label{eq_prob_bp_withouht_deadtime}\\
p_{(1)}(\mu_B^{eff}) &=& p_p (\mu_B^{eff}) + 2 p_d(\mu_B^{eff})\\
p_{(2)}(\mu_B^{eff}) &=& 2 p_{pd} (\mu_B^{eff}) + p_{dd} (\mu_B^{eff})\\
p_{(3)}(\mu_B^{eff}) &=& p_{pdd}(\mu_B^{eff})
\eqe

\subsection{Signal pulse detection probabilities}
Using the previous results, the probability that a detector registers a signal pulse, depending on the detector's polarization~$\phi$ and the signal pulse's polarization~$\theta$ with respect to the blinding pulse can be calculated giving
\8
p^S_{\theta, \phi} (\mu_B^{eff}, \mu_S^{eff}) =  (1-P_\phi(\mu_B^{eff})) P^S_\theta (\mu_S^{eff}) \quad ,
\9

with $\phi, \theta \in \{p,d,o\}$ meaning parallel, diagonal and orthogonal and $P_o(\mu_B^{eff})=0$ and the signal pulse mean photon number at the receiver $\mu_S^{eff}:=\eta_S \mu_S$ with mean photon number at the (signal) source $\mu_S$ and coupling efficiency from Alice to Bob $\eta_S$.

From this, the amount of information an adversary can gain from such an attack can be estimated:
The difference between the maximum ($=1$) and the current value of the binary entropy is used as the information gain a potential eavesdropper would have:
\eqa
I &=& 1- H_2(p(x_{Eve}=x_{Bob})) \\
&=& 1+ \Pr(x_{Eve}=x_{Bob})\log(\Pr(x_{Eve}=x_{Bob}))+\Pr(x_{Eve} \neq x_{Bob})\log(\Pr(x_{Eve} \neq x_{Bob}))\label{eq_inf_def}
\eqe

With
\eqa
p_\|(\mu_B^{eff}, \mu_S^{eff}) &:=& p^S_{p,p}(\mu_B^{eff}, \mu_S^{eff}) + p^S_{d,d}(\mu_B^{eff}, \mu_S^{eff}) \qquad \mbox{and} \\
p_\bot(\mu_B^{eff}, \mu_S^{eff}) &:=& p^S_{o,o}(\mu_B^{eff}, \mu_S^{eff}) + p^S_{d,d}(\mu_B^{eff}, \mu_S^{eff})
\eqe
it is intuitively clear that for large $\mu_B^{eff}$, i.e. high blinding intensities, all the terms with $p^S_{d,d}$ and $p^S_{p,p}$ will become small, because most of the time all detectors but the one orthogonal to the blinding pulse will be inactive.

Now the information gain (\ref{eq_inf_def}) can be calculated to give:
\eqa
I(\mu_B^{eff}, \mu_S^{eff}) &=& 1 + \frac{p_\|}{p_\| + p_\bot} \log_2 \left(\frac{p_\|}{p_\| + p_\bot}\right) + \frac{p_\bot}{p_\| + p_\bot} \log_2 \left(\frac{p_\bot}{p_\| + p_\bot}\right)
\eqe

In the simulation (Fig. \ref{fig_blink_theory}) it is assumed that the photon statistics of signal and blinding pulses  in a four SPAD receiver (Fig. 1) are Poissonian and thus $P_p(\mu_B^{eff}) = 1-e^{-\frac{\mu_B^{eff}}{2}}$, $P_d(\mu_B^{eff})=1-e^{-\frac{\mu_B^{eff}}{4}}$, $P^S_p(\mu_S^{eff}) = 1-e^{-\frac{\mu_S^{eff}}{2}}$ and $P^S_d(\mu_S^{eff}) = 1-e^{-\frac{\mu_S^{eff}}{4}}$.

\begin{acknowledgments}
This work was funded by the Elite Network of Bavaria program ``QCCC'' and the BMBF project ``QPENS''.

\end{acknowledgments}


\end{document}